\documentclass[twocolumn,aps,prl,groupedaddress,showpacs]{revtex4}
\usepackage{epsfig,amssymb,amsmath}
\def\comment#1{}\def\labell#1{\label{#1}}
\def\togli#1{}
\begin{document}
\fbox{{ {\sf\scriptsize Submitted draft}.}}
\title{Interferometric tunability of the absorption} \author{Vittorio
  Giovannetti$^1$, Seth Lloyd$^2$, Lorenzo Maccone$^3$}
\affiliation{$^{1}$NEST-CNR-INFM and Scuola Normale Superiore, Piazza
  dei Cavalieri 7, I-56126, Pisa, Italy.\\\vbox{$^{2}$MIT, Research
  Laboratory of Electronics and Dept. of Mech. Engin.,
  77  Mass. Av., Cambridge, MA 02139, USA.}\\
  \vbox{$^3$ Quantum Information Theory Group, Dip. Fisica ``A.
    Volta'', Pavia Univ., via Bassi 6, I-27100 Pavia, Italy.}}
\date{\today} 
\begin{abstract}
  We propose an interferometric setup that permits to tune the
  quantity of radiation absorbed by an object illuminated by a fixed
  light source. The method can be used to selectively irradiate
  portions of an object based on their transmissivities or to
  accurately estimate the transmissivities from rough absorption
  measurements.
\end{abstract}
\maketitle 

When an object is illuminated, it will absorb radiation proportionally
to its absorption coefficient: Darker portions of the object will
absorb more light than more transparent ones. Is there a way around
this? In this paper we analyze a setup which uses classical light
sources (i.e.  coherent beams) and permits to easily tune the quantity
of a light absorbed by an object independently on its transparency, by
appropriately tuning an interferometer phase. With the same setup,
high efficiency measurements of the absorption coefficient can be
performed via a feedback mechanism. The only underlying assumption is
that the object introduces a negligible dephasing into a probe beam.
Since we can employ quasi-monochromatic light, this assumption is met
in a variety of systems. Moreover, in the case of objects that have a
homogeneous phase image, the dephasing can be easily compensated with
the interferometer phase.

Our proposal draws inspiration from the so called 
``interaction-free-measurement'' setups,
\togli{\footnote{The ``interaction-free'' terminology is somewhat
  misleading, since the object does interact with the field.}}
 where a partially transparent object can be discriminated from a totally
transparent one with asymptotically negligible radiation
absorption~\cite{vaidman,qsd,imaging,kwiat,hradil}. Even though such
proposals were originally based on single-photon light pulses,
analogous results have been obtained also with classical
light~\cite{bjork,jang}.

The layout of the paper follows. We start by describing the proposed
interferometric setup. We show how the absorption peak can be tuned
and we analyze the irradiation selectivity. We then give the protocol
for high efficiency estimation of $\eta$. We conclude by analyzing
inhomogeneous objects, which incorporate different transmissivities.
Since the process does not involve any quantum effects (such as
entanglement or squeezing) one could also analyze it in terms of a
classical theory of radiation, instead of the quantum formalism we use
here for rigor.

\vspace{-.2cm}
\section{The apparatus}
\vspace{-.2cm}
The proposed apparatus is a modification of the experimental setup of
Ref.~\cite{jang}. It is obtained by concatenating a collection of $N$
Mach-Zehnder (MZ) interferometers and is depicted in
Fig.~\ref{f:schema}.  Initially a coherent state $|\alpha\rangle$
enters through one interferometer port (associated with the
annihilation operator $a_0$), and no photons enter from the other port
(associated with the annihilation operator $b_0$).  As shown in
Fig.~\ref{f:schema}, after each MZ, one of the two emerging beams (the
R beam) is focused on the object.  Then the two beams are recombined
at the input port of the next MZ.  After $N$ of such steps, the
radiation leaves the apparatus at the $N$th interferometer outputs
$a_N$ and $b_N$.  As we will show, appropriately tuning the
interferometers phase $\phi$ and the number $N$ of MZs it is possible
to choose the value of the transmissivity $\eta$ that will absorb the
most radiation in the object. The transmissivity $\eta$ of an object
is the probability that a single photon will pass through it or,
equivalently, the percentage of the transmitted intensity of an
impinging coherent beam. Note that an apparatus employing a {\em
  single} MZ which is crossed $N$ times by the light can also be
employed, where $N$ can be controlled by appropriately tilting one of
the interferometer mirrors~\cite{jang} or by using an acousto-optics
switch.
\begin{figure}[hbt]
\begin{center}
\epsfxsize=.9
\hsize\leavevmode\epsffile{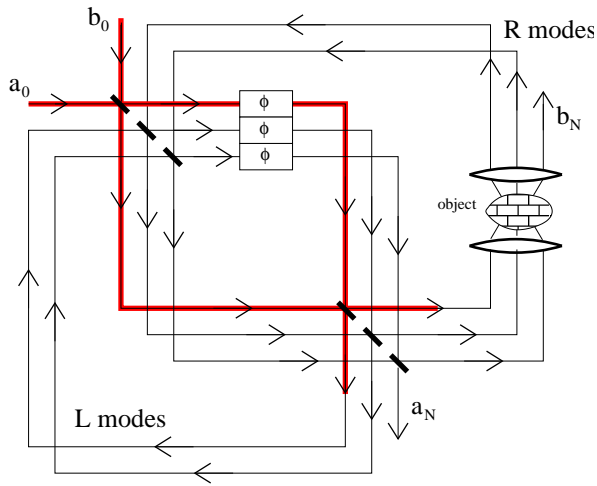}
\end{center}
\vspace{-.5cm}
\caption{Proposed apparatus.  It consists of $N$ 
  Mach-Zehnder (MZ) interferometers concatenated so that the output
  ports of the $n$th MZ is fed into the input ports of the successive
  one (for the sake of clarity the first interferometer is graphically
  enhanced).  All interferometers act on the radiation with the same
  phase shift $\phi$.  The object to be irradiated is placed outside
  the MZs and it interacts only with the R beams. Initially the
  radiation enters from the input $a_0$.  After $N$ round trips, it
  exits through the outputs $a_N$ and $b_N$.  }
\labell{f:schema}\end{figure}

The input-output relations of the interferometer can be obtained
observing that when two coherent states $|\alpha_n\rangle$ and
$|\beta_n\rangle$ impinge, respectively, into the input ports $a_n$
and $b_n$ of the $n$-th MZ (see Fig.~\ref{f:MZ}), the corresponding
outputs at ports $a_{n+1}$ and $b_{n+1}$ are still coherent states of
amplitudes $\alpha_{n+1}$ and $\beta_{n+1}$, given by
\begin{eqnarray}
\left(
\begin{array}{c}
\alpha_{n+1} \\
\beta_{n+1} \end{array}\right) = S \left(
\begin{array}{c}
\alpha_{n} \\
\beta_{n} \end{array}\right),
\end{eqnarray}
with
\begin{eqnarray}
S = e^{i\phi/2} 
\left( 
\begin{array}{cc}
\cos(\phi/2)  & i \sin(\phi/2)\\
 i \sin(\phi/2) & \cos(\phi/2) 
\end{array}
\right)\;,
\end{eqnarray}
where $\phi$ is the interferometer phase.
\begin{figure}[hbt]
\begin{center}
\epsfxsize=.9
\hsize\leavevmode\epsffile{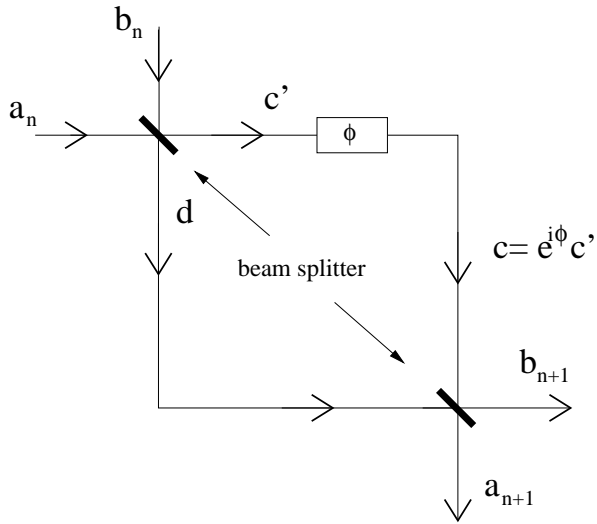}
\end{center}
\vspace{-.5cm}
\caption{\comment{Se serve spazio, si puo' eliminare questa figura}
  Mach-Zehnder interferometer constituting the $n$th element 
  in the Mach-Zehnder sequence in the apparatus of Fig~\ref{f:schema}.
  The first of the two 50-50 beam splitters transforms the input
  annihilation operators $a_n$, $b_n$ into
  $c^\prime=(a_n+b_n)/\sqrt{2}$ and $d=(b_n-a_n)/\sqrt{2}$
  respectively.  The second beam splitter transforms the annihilation
  operators $c$ and $d$ into $a_{n+1}= (c+d)/\sqrt{2}$ and $b_{n+1} =
  (d-c)/\sqrt{2}$.}  \labell{f:MZ}\end{figure}
The output $a_{n}$ is directly fed into a port of the MZ number $n+1$,
while the output $b_{n}$ first passes through the object and then
enters the other port of the same MZ.  Since the object absorbs each
photon with a probability $\eta$ without introducing any phase factor,
its action on the input coherent state $|\beta_{n+1}\rangle$ can be
modeled as a beam splitter with transmissivity $\eta$ that couples the
input radiation to a vacuum state and then discards one of the two
outputs. As a result the state $|\beta_{n+1}\rangle$ is transformed
into a coherent state of reduced amplitude $\sqrt{\eta}
\beta_{n+1}$~\cite{NOTA}. Thus, in the presence of the absorber, the
amplitudes $\alpha_{n+1}$ and $\beta_{n+1}$ of the coherent states at
the input of the MZ interferometer number $n+1$ is given by
\begin{eqnarray}
\left(
\begin{array}{c}
\alpha_{n+1} \\
\beta_{n+1} \end{array}\right) = S(\eta) \left(
\begin{array}{c}
\alpha_{n} \\ 
\beta_{n} \end{array}\right), \label{inout}
\end{eqnarray}
where 
\begin{eqnarray}
S(\eta) = e^{i\phi/2} 
\left( 
\begin{array}{cc}
\cos(\phi/2)  & i \sin(\phi/2)\\
i  \sqrt{\eta} \sin(\phi/2) & \sqrt{\eta} \cos(\phi/2) 
\end{array}
\right)\;. \label{mat}
\end{eqnarray}
Iterating Eq.~(\ref{inout}) $N$ times we can express the amplitude of
the coherent states emerging from the whole apparatus as
\begin{eqnarray}
\left(
\begin{array}{c}
\alpha_{N} \\
\beta_{N} \end{array}\right) = S^N(\eta) \left(
\begin{array}{c}
\alpha_{0} \\ 
0 \end{array}\right). \label{inout1}
\end{eqnarray}
Some examples of such evolution are given in Fig.~\ref{f:evol}, and an
analytic solution can be obtained by diagonalizing
$S(\eta)$~\cite{jang}.  The light absorbed by the object is given by
\begin{eqnarray}
I_{\mbox{\tiny ab}}= |\alpha_0|^2 - (|\alpha_N|^2 + |\beta_N|^2) \equiv
r \; |\alpha_0|^2\;,
\label{ass}
\end{eqnarray}
i.e. the input intensity $|\alpha_0|^2$ minus the total output
intensity $|\alpha_N|^2 + |\beta_N|^2$. The quantity $r$ is a
complicated function of $N$, $\phi$ and $\eta$ which can be explicitly
computed from Eq.~(\ref{inout1}). It measures the ``effective''
absorption constant of the object.

\vspace{-.2cm}
\section{Discussion} 
\vspace{-.2cm}
The possibility of changing the absorption of the illuminated object
from its natural value $1-\eta$ to an effective value $r\simeq 0$
allows one to determine the presence of a completely opaque object
(i.e.  $\eta=0$) with only an asymptotically small fraction of the
input radiation being absorbed~\cite{jang,imaging,kwiat,hradil,bjork}.
Consider Eq.~(\ref{inout1}) for $\eta=1$ (e.g.  completely transparent
object) and $\eta=0$ (e.g.  completely opaque object).  In these cases
simple analytical solutions can be obtained yielding
\begin{eqnarray}
\alpha_N &=& \alpha_0  e^{iN\phi/2} \cos(N \phi/2) \\
\beta_N &=& i\alpha_0 e^{iN\phi/2} \sin(N \phi/2) \label{sol1}\;,
\end{eqnarray}
for $\eta=1$, and 
\begin{eqnarray}
\alpha_N &=& \alpha_0  e^{iN\phi/2} \cos^N(\phi/2) \\
\beta_N &=& 0  \label{sol2}\;,
\end{eqnarray}
for $\eta=0$.  By choosing $\phi=\pi/N$, from Eqs.~(\ref{sol1})
and~(\ref{sol2}), it is immediate to see that all radiation exits from
the $b_N$-port if $\eta=1$ and that most of the radiation
(asymptotically all of it for $N\to\infty$) exits from the $a_N$-port
if $\eta=0$~\cite{NOTA1}. In both cases the light absorption is
minimal (i.e. exactly null in the first case and asymptotically null
in the second one).  Nonetheless they can be discriminated by simply
looking from which interferometer ports (e.g. $a_N$ or $b_N$) the
light emerges.

The possibility of controlling the effective absorption $r$ of the
object by changing the interferometer parameters is evident from
Fig.~\ref{f:evol2} where we plot $r$ as a function of the
transmissivity $\eta$ for different values of $\phi$ (choosing again
$N=\pi/\phi$): The function $r$ exhibits a peak for $\eta=\eta_{max}$
which increases from $\eta_{max} \sim 0$ to $\eta_{max}\sim 1$ as
$\phi$ decreases.  This effect can be explained intuitively as
follows.  For small values of $\phi$ (i.e.  high values of $N$) little
radiation is leaked into the R modes at every round trip with the
exception of the case when $\eta$ is high.  On the contrary, for small
values of $N$ (i.e. large values of $\phi$) a larger amount of
radiation is leaked into the R modes at every round trip, so that the
absorption peak moves to lower values of $\eta$. The dependence of the
absorption peak maximum as a function of $\phi$ and $N$ is depicted in
Fig.~\ref{f:peak}, left. This graph also shows the values of $\eta_{max}$
that can be attained in practice: it can be accurately fine-tuned only
for $\eta_{max}\gtrsim 0.5$ since only few discrete values of
$\eta_{max}$ are achievable for low $N$, whereas high $\eta_{max}\sim
1$ requires large $N$, which can be difficult to achieve practically.
Finally, the selectivity of the absorption, i.e.  the width of the
effective absorption curve $r$ as a function of $\eta$ (see
Fig.~\ref{f:evol2}), is not constant when $\phi$ is varied: The value
of the width-at-half-maximum is smaller for absorption curves peaked
at $\eta_{max} \simeq 0, 1$ and larger for $\eta_{max} \simeq 0.5$. In
the limit $\phi\to 0$ the $r$-curve becomes a very narrow spike peaked
just below $\eta=1$.

\begin{figure}[hbt]
\begin{center}
  \epsfxsize=.98\hsize\leavevmode\epsffile{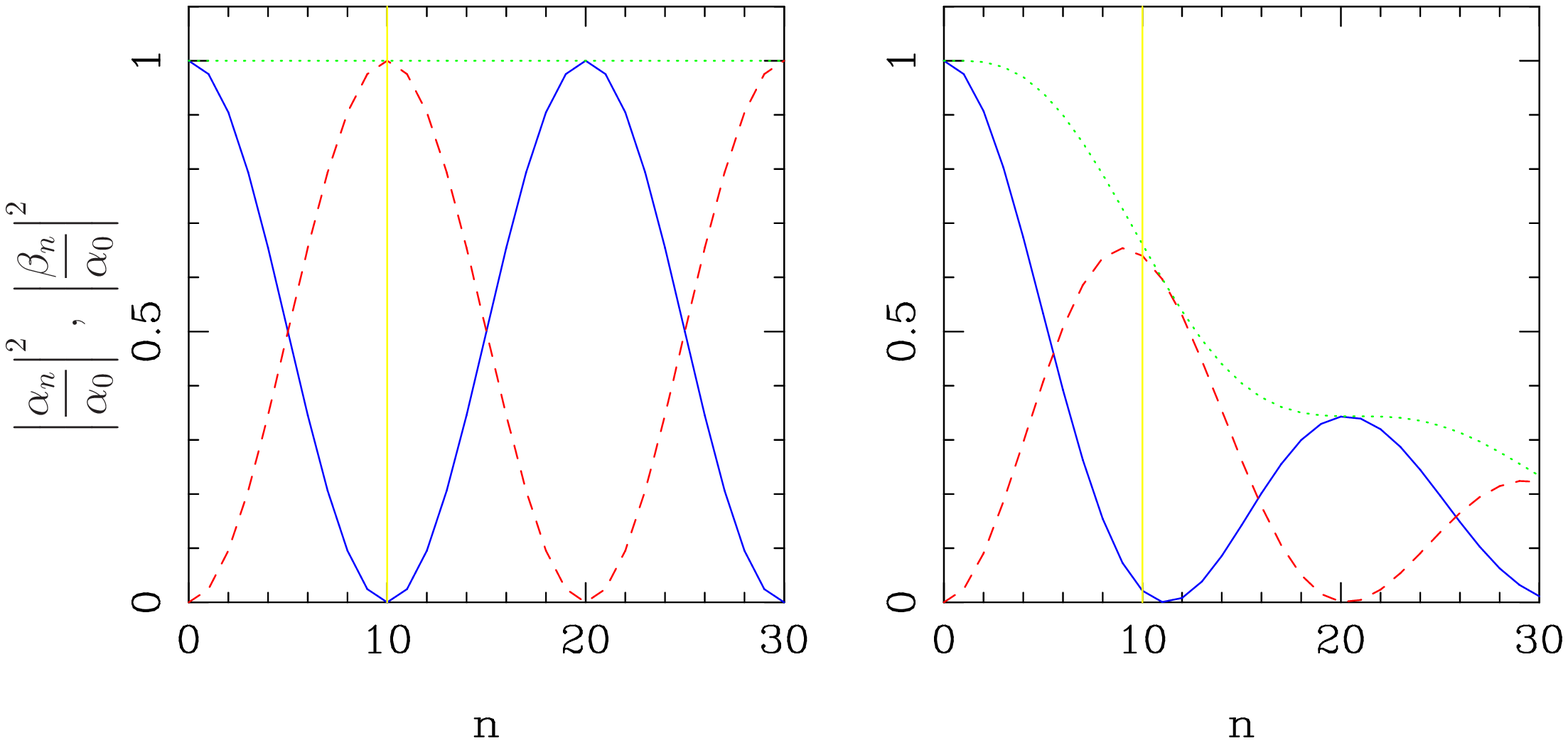}
\end{center}
\vspace{-.5cm}
\caption{\comment{Se serve spazio, si puo' eliminare questa figura}
  Plot of the (rescaled) output amplitudes of the $n$th MZ
  interferometer $|\alpha_n/\alpha_0|^2$ in the $a$-modes at L
  (continuous line) and $|\beta_n/\alpha_0|^2$ in the $b$-modes at R
  (dashed line). Initially all the radiation is in mode $a_0$, but, as
  the evolutions progresses, more and more radiation is transferred to
  the $b$-modes, until (for $n=\pi/\phi$) the radiation is entirely
  transferred.  Here $\phi=\pi/10$ so that the total transfer occurs
  for $n=10$ (vertical line). Left: The object is completely
  transparent ($\eta=1$), so that the total energy (dotted line) is
  constant; Right the object is semi-transparent ($\eta=.9$), so that
  the total energy decreases as the evolution progresses. }
\labell{f:evol}\end{figure}

\begin{figure}[hbt]
\begin{center}
  \epsfxsize=.48\hsize\leavevmode\epsffile{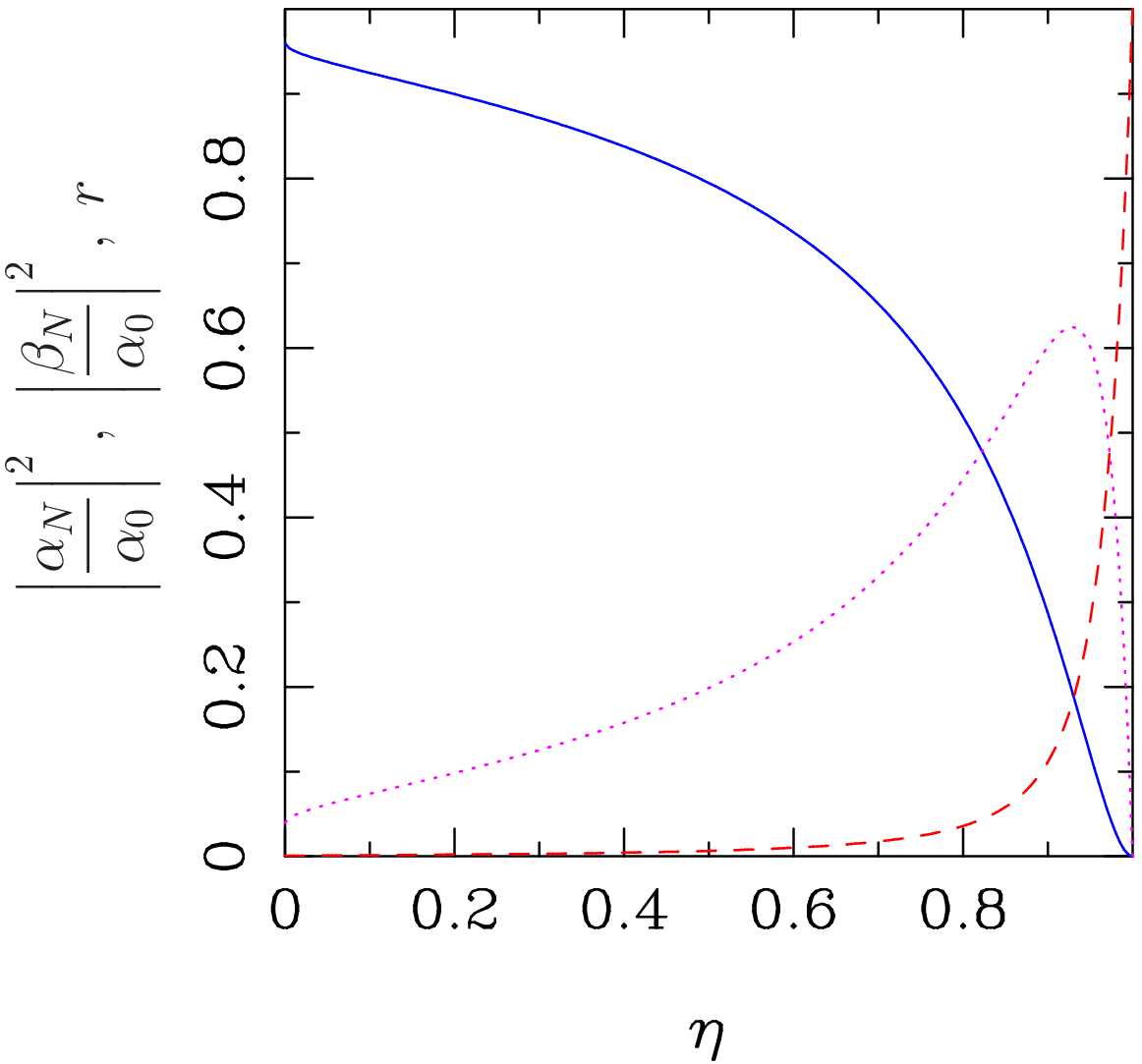}
  \epsfxsize=.46\hsize\leavevmode\epsffile{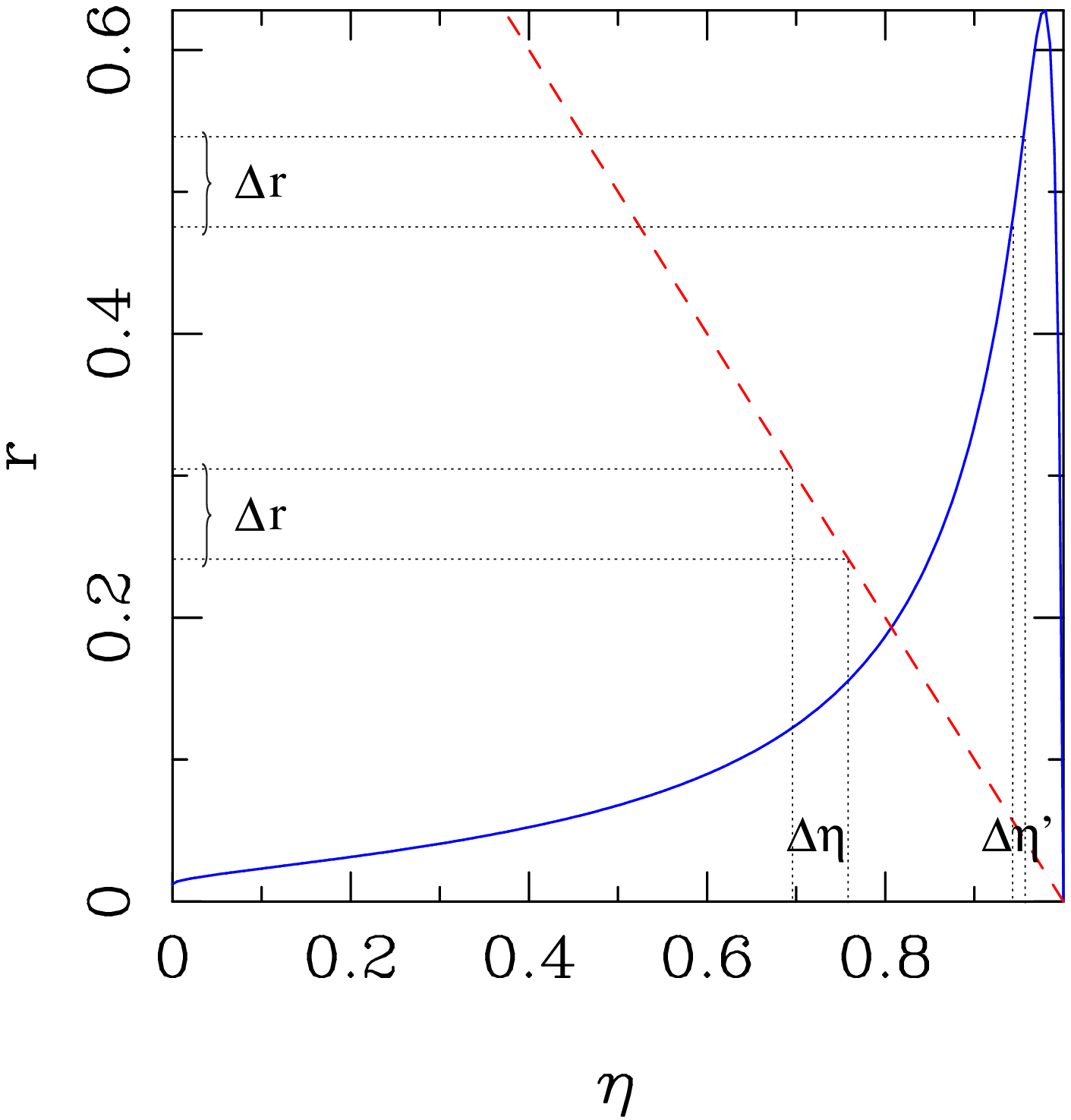}
\end{center}\vspace{-.4cm}\begin{center}
  \epsfxsize=.9\hsize\leavevmode\epsffile{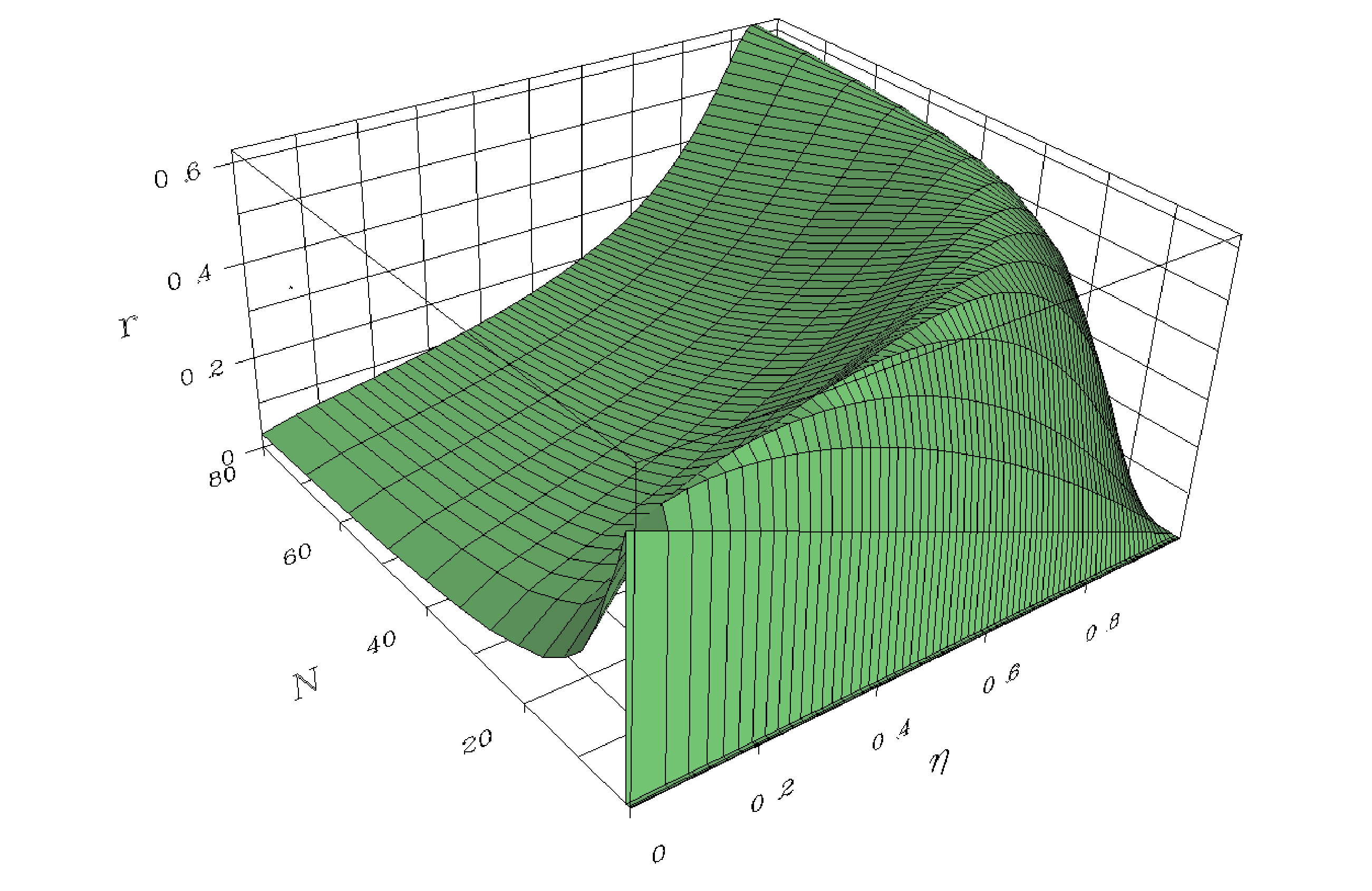}
\end{center}
\vspace{-.5cm}
\caption{Above left:~Plot of the (rescaled) apparatus output amplitudes
  $|\alpha_N/\alpha_0|^2$ at the output $a_N$ (continuous line) and
  $|\beta_N/\alpha_0|^2$ at $b_N$ (dashed line) as a function of the
  transmissivity $\eta$ of the object with $\phi=\pi/60$. For $\eta=1$
  (total transparency) all the output radiation is at $b_N$, whereas
  for $\eta=0$ (total absorption) all the output radiation is at $a_N$
  and a small amount of radiation has been absorbed.  The absorbed
  radiation is proportional to the effective absorption constant $r$,
  defined in Eq.~(\ref{ass}), which is depicted by the dotted lines.
  Above right: Plot of the absorption $r$ only, in the case
  $\phi=\pi/200$. This plot illustrates how the setup can be used to
  measure the absorption coefficient $\eta$ by measuring $r$,
  attaining a higher precision than with a direct measure (in which
  case the absorption is given by the dashed line): Starting from the
  same uncertainty $\Delta r$ in the measurement of $r$ with our
  procedure we can obtain a lower uncertainty $\Delta\eta'$ in the
  measurement of $\eta$ than the one, $\Delta\eta$, obtainable from a
  direct measurement. Below:~Plot of $r$ as a function of $\eta$ and
  $N=\pi/\phi$.  Notice how the absorption peak shifts as a function
  of $\phi$. The absorption peak moves to higher $\eta$ for decreasing
  $\phi$.} \labell{f:evol2}
\end{figure}

\begin{figure}[hbt]
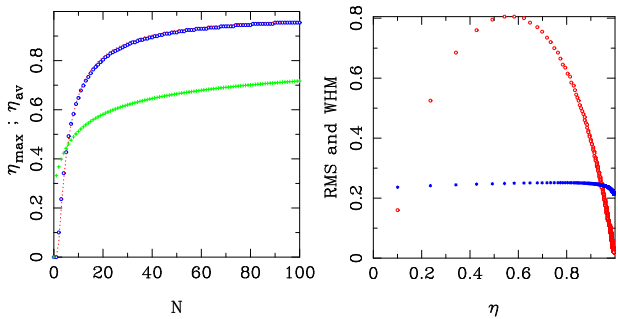

\begin{center}
  \epsfxsize=.48\hsize\leavevmode\epsffile{picco1.eps}
  \epsfxsize=.48\hsize\leavevmode\epsffile{picco2.eps}
\end{center}
\vspace{-.5cm}
\caption{Left:~Maximum  $\eta_{max}$ (circles) and
  average value $\eta_{av}$ (squares) of the absorption peak $r$ of
  Fig.~\ref{f:evol2} as a function of $N=\pi/\phi$.  (The maximum and
  the average follow different evolutions because of the asymmetry in
  the absorption curves). Increasing $N$ (i.e. decreasing $\phi$), the
  maximum in the absorption peak moves to higher values of $\eta$. The
  graph also details which are the actual values of $\eta$ that can be
  achieved through the proposed setup as a function of $N$. The dotted
  line is the function $[(N-1)/N]^4$ that gives a good interpolation
  of the peak evolution.  Right:~Selectivity in the irradiation as a
  function of the transmissivity peak. The width of the peaks of the
  dotted line in Fig.~\ref{f:evol2} is not uniform.  Here we plot the
  Root Mean Square (stars) and the width at half maximum (circles) of
  the absorption curve as a function of the absorption curve maximum.
  Notice that the RMS is almost constant over the whole range. }
\labell{f:peak}\end{figure}

\vspace{-.2cm}\section{High precision $\eta$-measurements}\vspace{-.2cm}
Our scheme can be easily adapted to high precision estimation of the
absorption coefficient, starting from low-quality measurements of the
absorption $r$. [The main idea is revealed by the lower right graph of
Fig.~\ref{f:evol2}.] The required iterative procedure is composed by
the following steps: i)~start by roughly estimating $\eta$ through an
absorption measurement and set the interferometer phase so that the
$r$ curve has a steep slope corresponding to such value of $\eta$;
ii)~perform another absorption measurement and estimate a better value
of $\eta$; iii)~again tune the interferometer phase, and so on. Since
the absorption curve $r$ for the values of $\eta\sim 1$ can be very
steep, a very good estimate of these $\eta$s can be achieved even when
the measurement of $r$ contains a large error $\Delta r$ (see
Fig.~\ref{f:evol2}).  Notice that the high values of $\eta\sim 1$ are
the hardest to estimate reliably without strongly irradiating the
object, since they are associated to the region of least transparency.

\vspace{-.2cm}\section{Inhomogeneous samples}\vspace{-.2cm}
In deriving Eq.~(\ref{inout}) we implicitly assumed that $\eta$ is
constant, i.e. that we employ a spatially homogeneous object with
uniform absorption within the waist of the light beams crossing it.
Instead, if it has spatially inhomogeneous absorption, we can still
use Eq.~(\ref{inout}) to describe the absorption of ``portions'' of
the incoming beam.  In fact, in the limit in which the scale of the
spatial inhomogeneities is much larger than the wavelength $\lambda$
of the source light, the diffraction of the propagating beam induced
by these inhomogeneities can be neglected.  In this regime the
illuminating beam can be effectively decomposed in independent
``sub-beams'' of the wavelength $\lambda$ and interacting
independently with the different portions of the object. A typical
example is when one performs imaging of a macroscopic object.
Controlling the interferometer parameters we can then selectively
choose which portion will absorb the most of the input radiation by
appropriately shifting the position of the absorption peak of $r$.

Explicitly, to selectively irradiate an inhomogeneous sample one thus
needs to: i)~estimate the transmissivity of the various portions of
the sample by conventional imaging techniques, identifying the
transmissivity $\bar\eta$ of the region that needs irradiation;
ii)~tune the phase $\phi=\pi/N$ so that the absorption is maximized
for $\bar\eta$.

\togli{\section{Conclusion} We have shown that the peak in the
  absorption curve $r$ of the interferometer of Fig.~\ref{f:schema}
  can be easily tuned by varying the interferometer parameters $N$ and
  $\phi$. In the above analysis we have assumed that the object does
  not introduce phase shifts to the radiation crossing it.  These
  would ultimately spoil the constructive interference in the
  Mach-Zehnder interferometers and prevent the build-up of the
  radiation in the right interferometer modes. \comment{aggiungere una
    spiegazione piu' dettagliata}. If, however, we know which phase
  $\theta$ is introduced, we can easily remove it by replacing the
  interferometer phase-shift $\phi$ with $\phi-\theta$.}

\par S. L. acknowledges financial support by ARDA, DARPA,
ARO, AFOSR, NSF, and CMI; V. G. was in part supported by the EC under
contract IST-SQUBIT2 and by the Quantum Information research program
of Centro di Ricerca Matematica Ennio De Giorgi of Scuola Normale
Superiore; L. M. acknowledges financial support by the Ministero
Italiano dell'Universit\`a e della Ricerca (MIUR) through FIRB (bando
2001) and PRIN 2005.


\begin{references}
\bibitem{NOTA}Rigorously speaking, the transformation $\beta_{n+1}
  \rightarrow \sqrt{\eta} \beta_{n+1}$ does not correspond to the
  linear mapping $|\beta_{n+1}\rangle \rightarrow |\sqrt{\eta}
  \beta{n+1}\rangle$ but to its density matrix counterpart:
  $|\beta_{n+1}\rangle \langle \beta_{n+1}| \rightarrow |\sqrt{\eta}
  \beta_{n+1}\rangle \langle \sqrt{\eta}\beta_{n+1}|$. The latter
  accounts for decoherence effect, whereas the former does not.  Since
  in our analysis we are always dealing with factorized states, the
  two transformations coincide for us.
\bibitem{NOTA1}This effect can be explained intuitively as follows. If
  the object is transparent, at the first round trip a small amount of
  radiation leaks into the R modes, at  the second round trip a higher
  amount  leaks   there   and so    on  constantly  increasing through
  constructive interference until all the   radiation moves into  such
  modes  after $N=\pi/\phi$ round  trips.  If, instead, the  object is
  opaque, the little radiation that has leaked into the R modes at the
  first  round  trip is   absorbed and does    not contribute  to  the
  constructive interference that would draw  more radiation into these
  modes at  the second round trip.  As a result very  little radiation
  transfers and most of it remains in the L modes.
\end{references}
 \end{document}